\documentclass[pra,showpacs,amsmath,amssymb,twocolumn]{revtex4}

\usepackage{amsthm}
\usepackage{dcolumn}
\usepackage{bm}

\newcommand\ket[1]{\ensuremath{|#1\rangle}}
\newcommand\bra[1]{\ensuremath{\langle#1|}}
\newcommand\iprod[2]{\ensuremath{\langle#1|#2\rangle}}
\newcommand\oprod[2]{\ensuremath{|#1\rangle\langle#2|}}
\newcommand\tr{\mathop{\rm tr}\nolimits}

\begin{document}

\title{Boundary effect of deterministic dense coding}

\author{Zhengfeng Ji}
  \email{jizhengfeng98@mails.tsinghua.edu.cn}
\author{Yuan Feng}
  \email{feng-y@tsinghua.edu.cn}
\author{Runyao Duan}
  \email{dry02@mails.tsinghua.edu.cn}
\author{Mingsheng Ying}
  \email{yingmsh@tsinghua.edu.cn}
\affiliation{
State Key Laboratory of Intelligent Technology and Systems, Department of Computer Science and Technology, Tsinghua University, Beijing 100084, China
}

\date{\today}

\begin{abstract}
We present a rigorous proof of an interesting boundary effect of deterministic dense coding first observed by Mozes et al. [Phys. Rev. A 71, 012311 (2005)]. Namely, it is shown that $d^2-1$ cannot be the maximal alphabet size of any isometric deterministic dense coding schemes utilizing $d$-level partial entanglement.
\end{abstract}


\maketitle

Dense coding~\cite{BW92} is a communication protocol that improves the capacity of a noiseless quantum channel with the assistance of quantum entanglement. The protocol, proposed by Bennett and Wiesner more than a decade ago, has now become one of the most important constituents of quantum information theory. Alice and Bob, two parties in the protocol, share in prior a maximally entangled pair and can communicate by sending a noiseless qubit. Without the entanglement, Alice can only transmit one of two different letters to Bob~\cite{YO93}. Surprisingly, however, she can do much better by utilizing the shared pair. Alice first performs an appropriate encoding operation on her half of the pair depending on the letter she wants to transmit and then sends the half on her side to Bob. Having the whole pair in hand, Bob is now able to perfectly distinguish these four possibilities as they form an orthonormal basis. Thus, by sending a single quantum bit, a classical transmission of one of four letters is achieved with the cost of consuming an entangled pair.

Illuminated by the original superdense coding protocol, many generalizations and related aspects have been considered in the literature. In Ref.~\cite{BW92}, the protocol is generalized to make use of maximally entangled $d$-level systems to transmit one letter out of $d^2$. Probabilistic and asymptotic approaches are taken in Refs.~\cite{HLG00,PPA04,WCSG05} and~\cite{BE95,HJS+96,BPV00,Hir01} respectively. Generalizations are also made to continuous variables~\cite{BK00,ZXP02} and to the multipartite cases~\cite{HLG01,LLTL02,BDL+04}. Recently, Mozes \textit{et~al.} initiated the discussion of deterministic dense coding~\cite{MOR05} using both numerical and analytical methods. We will give a mathematical proof of one of the interesting phenomena mentioned in their paper.

In deterministic dense coding, nonmaximal pure entanglement of two separated $d$-level systems is considered and we still want to reliably transmit one of the letters chosen from an alphabet. The main goal is to analyze the relation between the maximal size of the alphabet $N_{\max}(\psi)$ and the partially entangled state $\ket{\psi}$ in use. It is well-known that $N_{\max} = d^2$ for maximally entangled states and one might naturally expect to have $N_{\max}(\psi) = d^2-1$ for some $\ket{\psi}$ close enough to the maximal entanglement. Yet, results of numerical methods obtained in Ref.~\cite{MOR05} indicate that $N_{\max}(\psi)$ can be any value in the range $[d, d^2]$ with the possible exception of $d^2-1$. For $N_{\max} = d^2-2$ and $d=3,4$ the numerical algorithm can find solutions as desired while for $N_{\max} = d^2-1$ it fails definitely. However, as it is pointed out in Ref.~\cite{MOR05}, numerical methods cannot completely rule out the possibility of $d^2-1$, nor can it analyze the cases in higher dimensions. We will give a uniform proof of this observation for all $d\ge 2$ confirming that $N_{\max}(\psi) < d^2-1$ for any partial entanglement $\ket{\psi}$ of two $d$-level systems.

We now begin the proof by introducing some notations first. The partial entanglement $\ket{\psi}$ in use can be written in the following Schmidt decomposition~\cite{NC00} as
\begin{equation}
  \label{eq:schmidt}
  \ket{\psi} = \sum_{i=0}^{d-1} \sqrt{\lambda_i} \ket{i}_A \otimes \ket{i}_B,
\end{equation}
where $\lambda_i$ is a probability distribution and $\{ \ket{i}_A \}$ (resp. $\{ \ket{i}_B \}$) forms a basis of system $A$ (resp. $B$). Without loss of generality, we assume that $\lambda_i$ are already in descending order, that is,
\begin{equation*}
  \lambda_0 \ge \lambda_1 \ge \cdots \ge \lambda_{d-1}.
\end{equation*}

The main idea of deterministic dense coding is to encode classical messages by performing corresponding operations on system $A$ only, leaving states that can be perfectly identified on the whole system of $A$ and $B$. The most general operations that can be used here are quantum operations which is the case considered in Ref.~\cite{WCSG05}. We will focus on isometric encoding only in this paper and are thus interested in finding maximally sized set of local unitary operators $\{U_i^A\}_{i=0}^{N_{\max}(\psi)-1}$ such that $U_i^A\otimes I^B \ket{\psi}$ are orthogonal states.

If for some partial entanglement $\ket{\psi}$, $N_{\max}(\psi) = d^2-1$, we will see how this will lead to a contradiction. Let $\{V_i^A\}_{i=0}^{d^2-2}$ be the $d^2-1$ encoding operators. It is easy to see that applying an extra unitary operator afterward does not change the encoding efficiency, that is, $U_i^A = UV_i^A$ can be used equally in the dense coding protocol. We will specify unitary operator $U$ later. Denote $\ket{\psi_i} =  U_i^A\otimes I^B \ket{\psi}$, $\ket{\phi_{ij}} = U_i^A \ket{j}_A$. We have
\begin{equation*}
  \ket{\psi_i} = U_i^A\otimes I^B \ket{\psi} = \sum_{j=0}^{d-1} \sqrt{\lambda_j} \ket{\phi_{ij}} \ket{j}.
\end{equation*}
The orthogonality of $\ket{\psi_i}$ is equivalent to
\begin{eqnarray}
  \label{eq:orthogonal}
  \iprod{\psi_i}{\psi_j} & = & \bra{\psi} U_i^{A\dagger}U_j^A \otimes I^B \ket{\psi} \nonumber\\
    & = & \sum_{k,l} \sqrt{\lambda_k \lambda_l} \bra{k} U_i^{A\dagger}U_j^A \ket{l} \iprod{k}{l} \nonumber\\
    & = & \sum_k \lambda_k \bra{k}U_i^{A\dagger}U_j^A\ket{k} \nonumber\\
    & = & \tr \left( U_j^A \Lambda U_i^{A\dagger} \right) = \delta_{ij},
\end{eqnarray}
where $\Lambda$ is a $d\times d$ diagonal matrix of the Schmidt coefficients $\lambda_i$.

Let $P$ be the projector of the subspace spanned by $\{\ket{\psi_i},\,i=0,1,\ldots,d^2-2\} $,
\begin{equation}
  \label{eq:projector}
  P = \sum_{i=0}^{d^2-2}\oprod{\psi_i}{\psi_i} = \sum_{i,j,k} \sqrt{\lambda_j \lambda_k} \oprod{\phi_{ij}}{\phi_{ik}} \otimes \oprod{j}{k}.
\end{equation}
$Q=I-P$ is a projector of a one dimensional subspace and is also the density matrix of the pure state orthogonal to all $\ket{\psi_i}$.

We calculate the reduced density matrix $Q_A$ and $Q_B$ of $Q$ on system $A$ and $B$ respectively using Eq.~\eqref{eq:projector}.
\begin{eqnarray}
  \label{eq:Q_B}
  Q_B & = & \tr_A Q = \tr_A (I-P)\nonumber\\
      & = & dI - \sum_{i,j,k} \sqrt{\lambda_j \lambda_k} \iprod{\phi_{ik}}{\phi_{ij}} \oprod{j}{k} \nonumber\\
      & = & dI - \sum_{i,j} \lambda_j \oprod{j}{j} \nonumber\\
      & = & \sum_j \bigl(d-(d^2-1)\lambda_j\bigr) \oprod{j}{j},
\end{eqnarray}
where the fourth identity follows from the fact that $\{\ket{\phi_{ij}}\}_{j=0}^{d-1}$ forms an orthonormal basis of system $A$ for $i=0,1,\ldots,d^2-2$.
\begin{eqnarray}
  \label{eq:Q_A}
  Q_A & = & \tr_B Q = \tr_B (I-P)\nonumber\\
      & = & dI - \sum_{i,j,k}\sqrt{\lambda_j \lambda_k} \oprod{\phi_{ij}}{\phi_{ik}} \iprod{k}{j} \nonumber\\
      & = & dI - \sum_{i,j} \lambda_j \oprod{\phi_{ij}}{\phi_{ij}} \nonumber\\
      & = & dI - \sum_{i,j} \lambda_j U_i^A \oprod{j}{j} U_i^{A\dagger} \nonumber\\
      & = & \frac{1}{d^2-1} \sum_i U_i^A \biggl(\sum_j \bigl(d-(d^2-1)\lambda_j\bigr) \oprod{j}{j}\biggr) U_i^{A\dagger} \nonumber\\
      & = & \frac{1}{d^2-1} U \left( \sum_i V_i^A Q_B V_i^{A\dagger} \right) U^{\dagger}.
\end{eqnarray}

Since $Q$ is the density of a pure state, $Q_A$ and $Q_B$ have the same spectrum and we can choose $U$ properly such that $Q_A = Q_B$. Thus, we have
\begin{equation}
  Q_B = Q_A = \frac{1}{d^2-1} \sum_i U_i^A Q_B U_i^{A\dagger}.
\end{equation}

Let $\rho_i = U_i^A Q_B U_i^{A\dagger}$ and $p_i = 1/(d^2-1)$ for $i=0,1,\ldots,d^2-2$. It follows from the above equality that
\begin{equation*}
  S\left( \sum_i p_i \rho_i \right) = \sum_i p_i S(\rho_i) = S(Q_B),
\end{equation*}
where $S$ is the von Neumann entropy. This means that $\rho_i$ satisfy the equality condition of concavity of the entropy and are thus all equal. See section~11.3.5 of Ref.~\cite{NC00} for a detailed discussion of the concavity of von Neumann entropy and its equality condition. In our case, we have $U_i^A Q_B U_i^{A\dagger} = Q_B$ for all $i$, or equivalently
\begin{equation}
  \label{eq:commute}
  U_i^A Q_B = Q_B U_i^A.
\end{equation}

Eq.~\eqref{eq:Q_B} indicates that $Q_B$ is a diagonal matrix whose $(j, j)$-th element is $d-(d^2-1)\lambda_j$. Remember that $\ket{\psi}$ is a partial entanglement and thus not all $\lambda_j$ are equal. Suppose there are $t$ numbers of $\lambda_j$ having the same value as $\lambda_0$, then $1 \le t < d$ and
\begin{equation*}
  \lambda_0 = \cdots = \lambda_{t-1} > \lambda_t \ge \cdots \ge \lambda_{d-1}.
\end{equation*}
Thus, it follow from Eq.~\eqref{eq:commute} that the $(j,k)$-th element of matrix $U_i^A$ is $0$ for all $j < t \le k$ and $k < t \le j$. We can write $U_i^A = U_i^{A_0} \oplus U_i^{A_1}$ where $A_0$ and $A_1$ are subspaces spanned by $\{\ket{j}\}_{j=0}^{t-1}$ and $\{\ket{j}\}_{j=t}^{d-1}$. Denote $\mathcal{M} = M_t \oplus M_{d-t}$ where $M_n$ is the vector space of $n\times n$ matrices. Then $U_i^A$ is in $\mathcal{M}$ for all $i$ and the dimension of $\mathcal{M}$ is
\begin{equation}
  \label{eq:dimmax}
  \dim \mathcal{M} = t^2 + (d-t)^2 \le d^2 - 2d + 2.
\end{equation}

As $Q_B$ is a density matrix, each of its diagonal elements is less than or equal to $1$ and we have
\begin{equation}
  \label{eq:bound}
  \lambda_j \ge \frac{1}{d+1} > 0,\quad \text{for all } j = 0, 1, \ldots, d-1.
\end{equation}
Thus, for $M, N \in \mathcal{M}$, $\tr (M \Lambda N^{\dagger})$ defines an inner product of $M, N$ and makes $\mathcal{M}$ a Hilbert space. Eq.~\eqref{eq:orthogonal} indicates that $U_i^A$ are orthonormal vectors of space $\mathcal{M}$. However, the number of $U_i^A$, $d^2-1$, is strictly larger than the dimension of $\mathcal{M}$ which is at most $d^2-2d+2$. This contradicts with the basic facts of Hilbert spaces and it follows that it is impossible to find $d^2-1$ local unitary operators $V_i^A$ that transform $\ket{\psi}$ to orthogonal states. Namely, $N_{\max} \le d^2-2$ for all partial entangled states of two $d$-level systems.

As a summary, we have proved that the maximal alphabet size of any isometric dense coding schemes using a $d$-level entanglement cannot be $d^2-1$ no matter how close the partial entanglement is to the maximally entangled pair. In some sense, this boundary effect reveals the complex nature of deterministic dense coding. Although isometric deterministic dense encoding is the most natural and simplest form of generalization of the original dense coding process, it is yet not known whether unitary operators only are sufficient to fully utilize the partial entanglement. So whether general dense coding schemes can achieve an alphabet size of $d^2-1$ becomes an interesting problem for future research.

\section*{Acknowledgement}

We are thankful to the colleagues in the Quantum Computation and Information Research Group for helpful discussions. This work was supported by the Natural Science Foundation of China (Grant Nos. 60273003, 60433050 and 60305005).

\bibliography{d^2-1}

\end{document}